\documentclass[prl,aps,twocolumn,superscriptaddress]{revtex4-2}

\usepackage{amsmath,amssymb,amsfonts,bbm,color,graphicx}

\usepackage[utf8]{inputenc}
\usepackage[T1]{fontenc}
\usepackage{textcomp}

\usepackage[unicode=true,colorlinks=true]{hyperref}
\hypersetup{linkcolor=blue,citecolor=blue,urlcolor=blue}

\usepackage{bm}
\usepackage{dsfont}
\usepackage{xcolor}
\usepackage{txfonts} 
\usepackage{wasysym}

\usepackage{braket}

\newcommand{\iu}{\mathrm{i}} 
\newcommand{\eu}{\mathrm{e}} 
\newcommand{\du}{\mathrm{d}} 
\newcommand{\hc}{\mathrm{h.c.}} 

\newcommand{\bvec}[1]{\mathbf{#1}}

\newcommand{\Ztwo}{\mathbb{Z}_2}
\newcommand{\Uone}{\mathrm{U(1)}}
\newcommand{\SUtwo}{\mathrm{SU(2)}}

\newcommand{\SOthree}{\mathrm{SO(3)}}

\newcommand{\SO}{\mathrm{SO}}
\newcommand{\Hsq}{\mathcal{H}^{(2)}}
\newcommand{\Wsq}{{W}^{(2)}}
\newcommand{\Hhex}{\mathcal{H}^{(3)}}
\newcommand{\Whex}{{W}^{(3)}}
\newcommand{\Ch}{C} 
\newcommand{\Cm}{\nu_\text{M}}

\begin{document}

\title{Fractionalized fermionic quantum criticality in spin-orbital Mott insulators}

\author{Urban F. P. Seifert}
\affiliation{Institut f\"ur Theoretische Physik and W\"urzburg-Dresden Cluster of Excellence ct.qmat, Technische Universit\"at Dresden, 01062 Dresden, Germany}
\author{Xiao-Yu Dong}
\affiliation{Department of Physics and Astronomy, Ghent University, Krijgslaan 281, 9000 Gent, Belgium}
\author{Sreejith Chulliparambil}
\affiliation{Institut f\"ur Theoretische Physik and W\"urzburg-Dresden Cluster of Excellence ct.qmat, Technische Universit\"at Dresden, 01062 Dresden, Germany}
\affiliation{Max-Planck-Institut f{\"u}r Physik komplexer Systeme, N{\"o}thnitzer Stra{\ss}e 38, 01187 Dresden, Germany}
\author{Matthias Vojta}
\affiliation{Institut f\"ur Theoretische Physik and W\"urzburg-Dresden Cluster of Excellence ct.qmat, Technische Universit\"at Dresden, 01062 Dresden, Germany}
\author{Hong-Hao Tu}
\affiliation{Institut f\"ur Theoretische Physik and W\"urzburg-Dresden Cluster of Excellence ct.qmat, Technische Universit\"at Dresden, 01062 Dresden, Germany}
\author{Lukas Janssen}
\affiliation{Institut f\"ur Theoretische Physik and W\"urzburg-Dresden Cluster of Excellence ct.qmat, Technische Universit\"at Dresden, 01062 Dresden, Germany}

\begin{abstract}
We study transitions between topological phases featuring emergent fractionalized excitations in two-dimensional models for Mott insulators with spin and orbital degrees of freedom. 
The models realize fermionic quantum critical points in fractionalized Gross-Neveu* universality classes in (2+1) dimensions. They are characterized by the same set of critical exponents as their ordinary Gross-Neveu counterparts, but feature a different energy spectrum, reflecting the nontrivial topology of the adjacent phases.
We exemplify this in a square-lattice model, for which an exact mapping to a $t$-$V$ model of spinless fermions allows us to make use of large-scale numerical results, as well as in a honeycomb-lattice model, for which we employ $\epsilon$-expansion and large-$N$ methods to estimate the critical behavior.
Our results are potentially relevant for Mott insulators with $d^1$ electronic configurations and strong spin-orbit coupling, or for twisted bilayer structures of Kitaev materials.
\end{abstract}

\date{\today}

\maketitle

Topology has been established as an organizing principle for states of matter beyond the Landau paradigm of symmetry-breaking.
Significant progress has been made in systematically understanding symmetry-protected topological (SPT) phases \cite{chen13,senthil15}, in particular in one dimension \cite{pollmann10,chen11,schuch11,turner11,kitaev11}, the study of their phase transitions \cite{grover13,lu14,slagle15}, and the nature of the respective critical points \cite{verresen17,qin17}.
In contrast to these short-range entangled SPT states, universal properties of phases with intrinsic topological order are less well understood.
Their long-range entanglement structure leads to highly unconventional features, such as emergent deconfined gauge fields and fractionalized excitations
\footnote{Here, we refer to a phase as being topological if its low-energy excitations belong to superselection sectors that are invariant under the action of local operators \cite{bonderson13,sachdev18}. This notion thus also includes gapless $\Ztwo$ and $\Uone$ spin liquids.}.

While conventional phase transitions can be understood by analyzing the fluctuations of a local order parameter in a Landau-Ginzburg-Wilson framework, the absence of such an order parameter in topological phases raises fundamental questions: What kind of (continuous) transitions involving topological phases are possible, and what are the universal properties of these unconventional quantum critical points \cite{xu12}?
Typically, systems hosting topological order involve strong interactions, so that only few controlled analytical studies and numerical results of the corresponding transitions are available, and are mostly limited to toy models \cite{vojta18}.
An important example is given by a model of hard-core bosons, which has been shown to feature a fractionalized quantum critical point in the (2+1)-dimensional XY* universality class \cite{isakov12}. This unconventional universality class differs from the ordinary XY universality as a consequence of the topological degeneracy and the fact that only states with even numbers of fractionalized particles are allowed in the spectrum.
Similarly, fractionalized counterparts of the ordinary Ising and O($N$) universality classes have been discussed \cite{whitsitt16, schuler16}.
An effective model that realizes a related transition in the presence of gapless fermions has also been proposed \cite{gazit18}.
However, although a number of topological phases are characterized by emergent fermionic excitations \cite{rantner01, kitaev06, senthil08, he17, song19}, a microscopic model that exhibits a fractionalized version of a fermionic quantum critical point appears as yet unknown.

In this Letter, we construct two such examples.
Specifically, we show that in spin-orbital models featuring emergent gapless Majorana excitations coupled to a $\Ztwo$ gauge field, there are continuous quantum phase transitions across which a global $\Ztwo$ or $\SOthree$ spin rotation symmetry is spontaneously broken and (a subset of) the Majorana fermions become gapped out. These fractionalized quantum critical points fall into fermionic Gross-Neveu* universality classes in (2+1) dimensions, the nontopological counterparts of which have arouse significant interest lately in the context of interacting Dirac fermion systems \cite{*[] [{; and references therein.}] boyack20}.

{\em Model construction.---}%
Our starting points are spin-orbital implementations \cite{yzk09,yaolee11,nakai12,deCarva18,natori20} of the bond-dependent Kitaev exchange interaction \cite{kitaev06}, which belong to a family of exactly soluble generalized Kitaev models recently introduced \cite{CSVJT20}.
These models have quantum spin-orbital-liquid ground states with static gapped $\Ztwo$-vortex excitations and $\Cm$
itinerant gapless Majorana fermions hopping on the square ($\Cm$ even) or honeycomb ($\Cm$ odd) lattices. Adding three-body interactions induces chiral next-nearest-neighbor hopping of the Majorana fermions and opens up a topologically non-trivial band gap with Chern number $\Ch=\Cm$, giving rise to the sixteen anyon theories as classified in Kitaev's sixteenfold way \cite{kitaev06}.

We exploit the fact that these spin-orbital Kitaev models allow for simple antiferromagnetic Heisenberg (and Ising, respectively) spin interactions, which leave the vortex excitations static.
Considering Mott insulators with orbital degeneracy and sizable bond-dependent exchanges, such perturbations are expected to be present in the respective material-specific Kugel-Khomskii \cite{kk77,kk82} models.
Adding the perturbations spoils the exact solubility, but the $\Ztwo$ fluxes carried by the vortices remain good quantum numbers, such that resulting transitions are driven purely by interactions among the itinerant Majorana fermions.
This allows us to find controlled theoretical descriptions based on the fermionic parton construction of the unperturbed model, and to use already available high-precision numerical results and established analytical techniques to study the resulting problems of interacting Dirac fermions.

\begin{figure}[tb]
\includegraphics[width=\columnwidth]{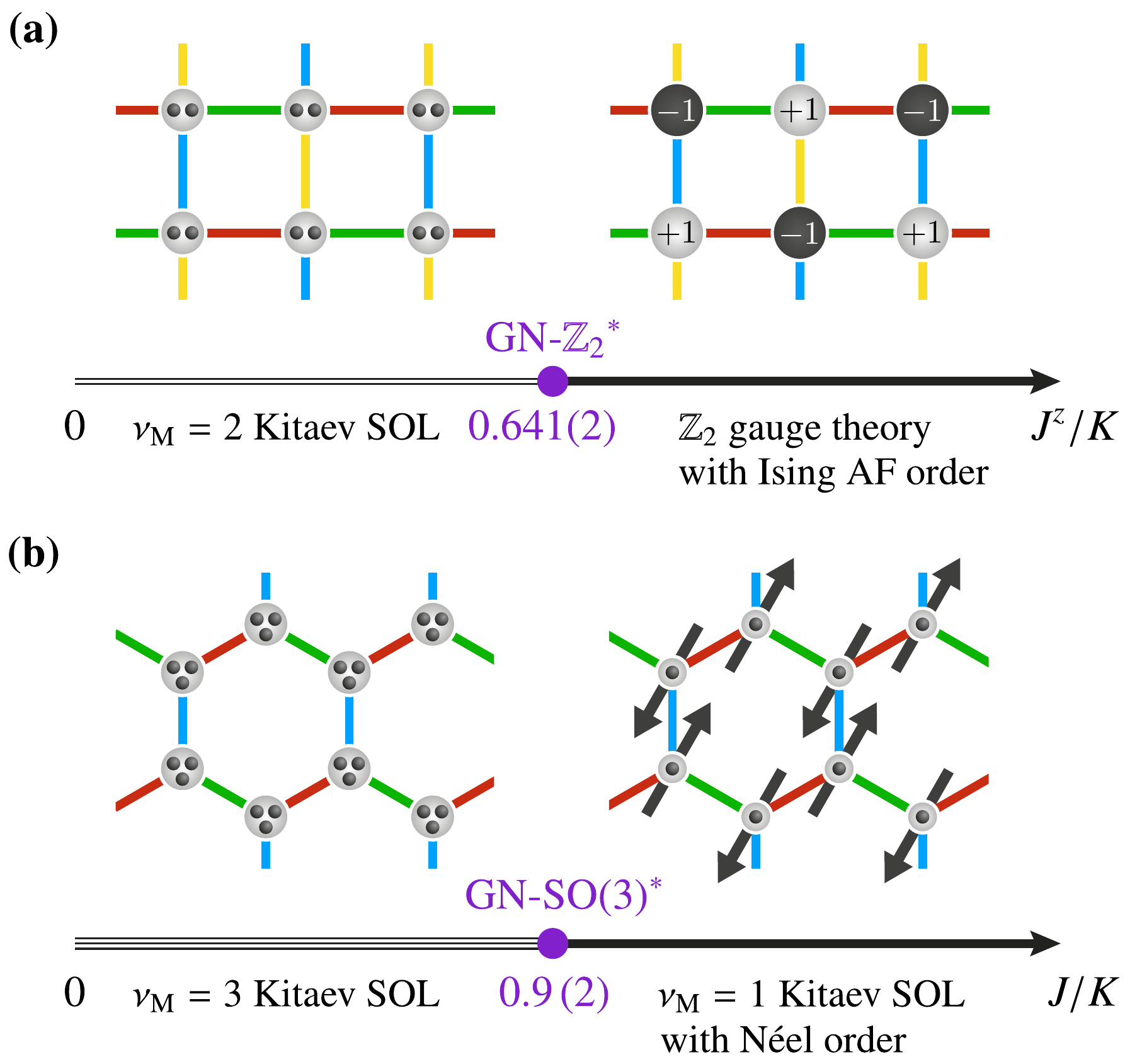}
\caption{(a) Adding a sufficiently strong antiferromagnetic Ising spin interaction to the $\Cm=2$ Kitaev spin-orbital liquid (SOL) on the square lattice gives rise to Ising antiferromagnetic order in the spin sector, with the remaining degrees of freedom described in terms of a $\mathbb{Z}_2$ gauge theory.
The continuous transition at $J_\mathrm{c}^z/K = 0.641(2)$ is in the Gross-Neveu-$\mathbb{Z}_2$* universality class.
(b) The honeycomb-lattice model features a Gross-Neveu-$\SO(3)$* quantum critical point at $J_\mathrm{c}/K = 0.9(2)$ between the $\Cm=3$ Kitaev SOL and a $\Cm=1$ Kitaev SOL with Néel order in the spin sector.
}
\label{fig:introfig}
\end{figure}

{\em Gross-Neveu* transitions.---}%
For the $\Cm=2$ spin-orbital model on the square lattice with an Ising perturbation, we find an exact mapping to an interacting fermion-hopping problem on the $\pi$-flux lattice.
This fermionic model has been studied before using large-scale quantum Monte Carlo simulations \cite{wang14,li15,huffman17,huffman20}, allowing us to determine the phase diagram to high accuracy. Translating these results back to the spin-orbital model reveals the existence of a fractionalized fermionic quantum critical point in the Gross-Neveu-$\Ztwo$* universality class, separating a gapless $\Ztwo$-symmetric spin-orbital liquid from a partially ordered and fully gapped phase with Ising antiferromagnetic order in the spin sector, see Fig.~\ref{fig:introfig}(a).
We furthermore show that the $\Cm=3$ spin-orbital model on the honeycomb lattice with a Heisenberg perturbation in the spin sector features a similar fermionic quantum critical point between a gapless $\SO(3)$-symmetric spin-orbital liquid and a partially ordered and partially gapped phase, in which the $\SO(3)$ symmetry is spontaneously broken, Fig.~\ref{fig:introfig}(b).
The corresponding universality class, dubbed Gross-Neveu-$\SO(3)$*, is a fractionalized version of a new member of the Gross-Neveu family, and we determine its critical behavior below.

Similar to the fractionalized bosonic transitions~\cite{schuler16}, the Gross-Neveu* universality classes are characterized by a spectrum that reflects the topological degeneracy of the adjacent phases and the additional constraints on the physical states [\onlinecite{suppl}\textcolor{blue}{(a)}].
However, here the order parameters are composite operators consisting of pairs of fractionalized particles and are therefore gauge invariant, and numerically and experimentally accessible.
This means that not only the correlation-length exponents $\nu$, but also the order-parameter anomalous dimensions $\eta_\phi$ in the Gross-Neveu and Gross-Neveu* universality classes coincide, in contrast to the bosonic situation \cite{isakov12}. 
We emphasize that this is not the case for the fermionic anomalous dimension $\eta_\psi$, which has a physical meaning only in the ordinary Gross-Neveu universality classes.
Further, the change of the number of gapless Majorana fermions at the transition can be understood as a signature of distinct topological orders in the two adjacent phases \footnote{We may take the viewpoint that the two respective phases connected by the transition have parent gapped topologically ordered phases obtained by opening a topologically non-trivial gap for each itinerant Majorana fermion. The topological data of the two respective low-energy anyon theories of the gapped topological phases as classified in Kitaev's sixteenfold way \cite{kitaev06} is then found to differ.}.
Lastly, at elevated temperatures, additional nontrivial physics can be expected as a consequence of thermal flux excitations \cite{motome15}.

{\em Ising order on the square lattice.}---%
The $\Cm = 2$ spin-orbital liquid is described by a Hamiltonian on the square lattice with a two-site unit cell and a biquadratic XY-spin and Kitaev-type-orbital interaction,
\begin{equation} \label{eq:h2}
    \Hsq_K = - K \sum_{\langle ij \rangle_\gamma} \left( \sigma^x_i \sigma^x_j + \sigma^y_i \sigma^y_j \right) \otimes \tau^\gamma_i \tau^\gamma_j, \qquad K > 0,
\end{equation}
where the Pauli matrices $(\sigma^x,\sigma^y)$ [$(\tau^\gamma) = (\tau^x, \tau^y, \tau^z, \mathds{1})$] act on spin (orbital) degrees of freedom, and $\langle ij \rangle_\gamma$, $\gamma = 1, \dots, 4$, denote the four inequivalent bonds in the unit cell. 
Representing the spin-orbital degrees of freedom using Majorana fermions [\onlinecite{suppl}\textcolor{blue}{(b)}], Eq.~\eqref{eq:h2} can be mapped to a problem of two dispersing Majorana fermions $c^x,c^y$ in the background of a $\Ztwo$ gauge field \cite{kitaev06,CSVJT20}, $\Hsq_K \mapsto K \sum_{\langle ij \rangle} \iu u_{ij} ( c^x_i c^x_j + c^y_i c^y_j )$, with $i \in A$, $j\in B$ sublattice.
Importantly, Eq.~\eqref{eq:h2} possesses an extensive number of conserved quantities given by the two (symmetry-inequivalent) plaquette operators $\Wsq_p = \sigma_k^z \sigma_n^z \otimes \tau^x_i \tau^y_j \tau^x_k \tau^y_n$ and $\Wsq_{p'} = \sigma^z_k \sigma^z_n \otimes \tau^y_k \tau^x_l \tau^y_m \tau^x_n$,
which correspond to elementary Wilson loop operators for the gauge field $u_{ij}$ and thus constrain fluxes as excitations of the gauge field to be \emph{static}.
By Lieb's theorem \cite{lieb94}, the ground-state flux configuration is given by $\Wsq_p = \Wsq_{p'} = -1$ for all $p$, $p'$.

We now add to $\Hsq_{K}$ an antiferromagnetic Ising spin interaction of the form
\begin{align}
    \Hsq_{J^z} = J^z \sum_{\langle i j \rangle} \sigma^z_i \sigma^z_j \otimes \mathds{1}_i \mathds{1}_j, \qquad J^z >0.
\end{align}
Importantly, from $[\Hsq_{J^z}, \Wsq_p] = [\Hsq_{J^z}, \Wsq_{p'}] = 0$ it follows that $\Ztwo$ gauge fluxes remain static in the full system $\Hsq = \Hsq_{K} +\Hsq_{J^z}$.
Note that the exact solubility is spoiled at finite $J^z$, since upon mapping to Majorana fermions, the Ising term introduces short-range interactions, $\Hsq_{J^z} \mapsto - J^z \sum_{\langle ij \rangle} c^x_i c^y_i c^x_j c^y_j $.
Using complex fermions $f_j = (c^x_j + \iu c^y_j)/2$, the problem maps to
\begin{equation} \label{eq:spinlessInt}
    \Hsq \mapsto \sum_{\langle ij \rangle} \left[ 2 K u_{ij} \left(f_i^\dagger f_j + f_j^\dagger f_i\right) + 4J^z \left(n_i - \tfrac12 \right) \left(n_j - \tfrac12 \right) \right],
\end{equation}
where $n_j = f_j^\dagger f_j$ is the fermion density operator and we have further performed a gauge transformation $f_j \mapsto -\iu f_j$ for $j \in B$.
Note that the $\SO(2)$ spin rotation symmetry now corresponds to a global $\Uone$ phase rotation symmetry.
Lieb's theorem remains applicable for finite $J^z$ \cite{lieb94,nachtergaele96}, and thus the ground-state flux sector of $\Hsq$ describes a tight-binding model of spinless fermions on the $\pi$-flux lattice with hopping parameter $t \equiv 2K$ and nearest-neighbor repulsion $V \equiv 4J^z$ at half filling.
In the noninteracting limit $J^z \ll K$, the spectrum features two Dirac nodes at the Fermi level, describing a Dirac semimetal phase in the fermionic model and a quantum paramagnet in the original theory.
In the strong-coupling limit $J^z\gg K$, the system favors a charge-density-wave (CDW) state, in which the $\Ztwo$ order parameter $\rho = (n_{i,A}-n_{j,B})/2$, where $n_{i,A}$ ($n_{j,B}$) refers to the fermion density on the $A$ ($B$) sublattice, acquires a finite expectation value.
In the spin-orbital basis, the CDW state corresponds to Ising antiferromagnetic order in the spin sector, $\langle \rho \rangle = \frac14 \langle\sigma_{i,A}^z-\sigma_{j,B}^z \rangle \neq 0$, while the orbital degrees of freedom feature a $\Ztwo$ gauge structure [\onlinecite{suppl}\textcolor{blue}{(c)}].
Because Dirac fermions are stable against weak perturbations, we expect the order-disorder transition to occur at finite $J^z/K$.
In fact, the fermionic model on the $\pi$-flux lattice has been studied before using large-scale quantum Monte Carlo simulations \cite{wang14,li15,huffman17}, which show a single continuous transition at $J^z_\mathrm{c} = 0.641(2) K$ \cite{huffman20}, characterized by the critical exponents $\eta_\phi = 0.51(3)$, $1/\nu = 1.12(1)$, and dynamical exponent $z=1$.
The quantum critical point in the fermionic model falls into the (2+1)-dimensional Gross-Neveu-$\Ztwo$ universality, which has been excessively investigated in recent years~%
\cite{%
mihaila17,zerf17,ihrig18,
gracey16,gracey17,
iliesiu16,iliesiu18,
braun11,janssen14,vacca15,knorr16,gies17,dabelow19,
chandrasekharan13,
wang15,wang16,hesselmann16,
he18,chen19,schuler19,liu20
}.
Consequently, the transition in the spin-orbital model falls into the Gross-Neveu-$\Ztwo$* universality class and is characterized by the same universal exponents.

{\em N\'{e}el antiferromagnet on the honeycomb lattice.}---%
On the honeycomb lattice, a spin-orbital liquid can be stabilized in a model with a biquadratic Heisenberg-spin and Kitaev-orbital interaction \cite{yaolee11,CSVJT20},
\begin{equation} \label{eq:h3}
    \Hhex_K = -K \sum_{\langle ij \rangle_\gamma} \vec \sigma_i \cdot \vec \sigma_j \otimes \tau^\gamma_i \tau^\gamma_j, \qquad K > 0,
\end{equation}
where now $\langle i j\rangle_{\gamma}$, $\gamma = 1,2,3$, refer to the three inequivalent bonds in the two-site unit cell and $\vec\sigma = (\sigma^x,\sigma^y,\sigma^z)$.
As before, the spin-orbital operators can be represented by Majorana fermions, leading to a problem of three dispersing Majorana fermions $c_i = (c_i^x,c_i^y,c_i^z)^\top$ coupled to a $\Ztwo$ gauge field $u_{ij}$, $\Hhex_K \mapsto K \sum_{\langle ij \rangle} \iu u_{ij} c^\top_i c_j$.
The gauge field is static as a result of the conservation of the flux operators $\Whex_p = \mathds{1} \otimes \tau^x_i \tau^y_j \tau^z_k \tau^x_l \tau^y_m \tau^z_n$.
The ground state of $\Hhex_K$ lies in the flux-free sector $\Whex_p = +1$ for all $p$ \cite{lieb94}, and the three Majorana fermions lead to a well-defined spectrum on one half of the lattice's Brillouin zone, featuring one complex Dirac node per Majorana flavor.
Note that $\Hhex_K$ possesses a global symmetry under $\SO(3)$ spin rotations, which in the fermionic representation corresponds to a flavor rotation.

We now add an antiferromagnetic Heisenberg interaction among only the spin degrees of freedom of the form
\begin{equation} \label{eq:hj}
    \Hhex_J = J \sum_{\langle i j \rangle} \vec \sigma_i \cdot \vec \sigma_j \otimes \mathds{1}_i \mathds{1}_j, \qquad J>0.
\end{equation}
Crucially, the flux operators remain static since $[\Hhex_J,\Whex_p] = 0$.
Such a spin-only Heisenberg interaction occurs generically in spin-orbital systems due to orbital-diagonal superexchange interactions \cite{kk82,natori19}.
Mapping $\Hhex_J$ to the Majorana representation yields 
$\Hhex_J \mapsto \frac{J}{4}  \sum_{\langle ij \rangle} (c_i^\top \vec L c_i) \cdot (c_j^\top \vec L c_j)$, with the $\SO(3)$ generators $L^\alpha_{\beta \gamma} = - \iu \epsilon^{\alpha \beta \gamma}$ in the fundamental representation,
revealing that $\Hhex_J$ again maps to short-range interactions.

For $J\ll K$, the ground state of $\Hhex = \Hhex_K + \Hhex_J$ is a semimetal with three flavors of gapless Dirac excitations, corresponding to the $\Cm=3$ spin-orbital liquid.
For $J\gg K$, we expect the vector order parameter $\vec n = (c^\top_{i,A} \vec L c_{i,A} - c^\top_{j,B} \vec L c_{j,B})/4$ to acquire a finite expectation value, e.g., $\langle \vec n \rangle \propto \hat z$ without loss of generality.
This breaks the $\SO(3)$ symmetry to a residual $\SO(2)\times \Ztwo$ symmetry and gaps out two of the three Majoranas. However, since $L^z$ has a zero eigenvalue, the third Majorana mode remains gapless in the ordered phase.
In the spin-orbital basis, we have $\langle\vec n\rangle = \langle \vec \sigma_{i,A} - \vec \sigma_{j,B} \rangle / 2$. The symmetry-broken phase thus corresponds to N\'{e}el antiferromagnetic order in the spin sector.
This phase can also be understood within a simple mean-field decoupling $\vec \sigma_i \simeq \langle \vec \sigma_i \rangle$ in the spin-orbital formulation of the model, yielding
\begin{equation} \label{eq:spin-mean-field}
    \Hhex_K + \Hhex_J \simeq K |\langle\vec n\rangle|^2 \sum_{\langle i j \rangle_\gamma} \tau^\gamma_i \tau^\gamma_j -  \frac{3 N_\mathrm{uc} J}{2} |\langle \vec n\rangle|^2,
\end{equation}
where $N_{\mathrm{uc}}$ is the number of unit cells. For large $J\gg K$, the spins order antiferromagnetically. The remaining orbital degrees of freedom are described by the $\Cm=1$ Kitaev honeycomb model with an effective antiferromagnetic Kitaev coupling $K |\langle\vec n\rangle|^2$.

To investigate the model at finite $J/K$, we employ Majorana mean-field theory. We decouple into onsite fields $\langle \vec\sigma_i \rangle = \langle c^\top_i \vec L c_i\rangle/2$, transforming as vectors under $\SO(3)$, and singlet bond variables $\chi_{ij} = \langle \iu c^\top_i c_j \rangle / 3$.
In the limiting cases for weak and strong interactions, the ground state is flux free \cite{lieb94}. Assuming that this property holds also at intermediate values of $J/K$, and restricting ourselves to isotropic and translation-invariant mean fields, we can solve the self-consistency equations iteratively [\onlinecite{suppl}\textcolor{blue}{(d)}].
At $J_\mathrm{c} \approx 0.6 K$, we find a direct continuous transition from the $\Cm=3$ spin-orbital liquid to the N\'{e}el-ordered phase with a single gapless itinerant Majorana fermion, corresponding to $\Cm = 1$.
We note that the true $J_\mathrm{c}/K$ should be expected to be larger than the mean-field result, as quantum fluctuations tend to destabilize the antiferromagnetic order~\cite{sorella92, assaad13}.

\begin{figure}
\includegraphics[width=\columnwidth]{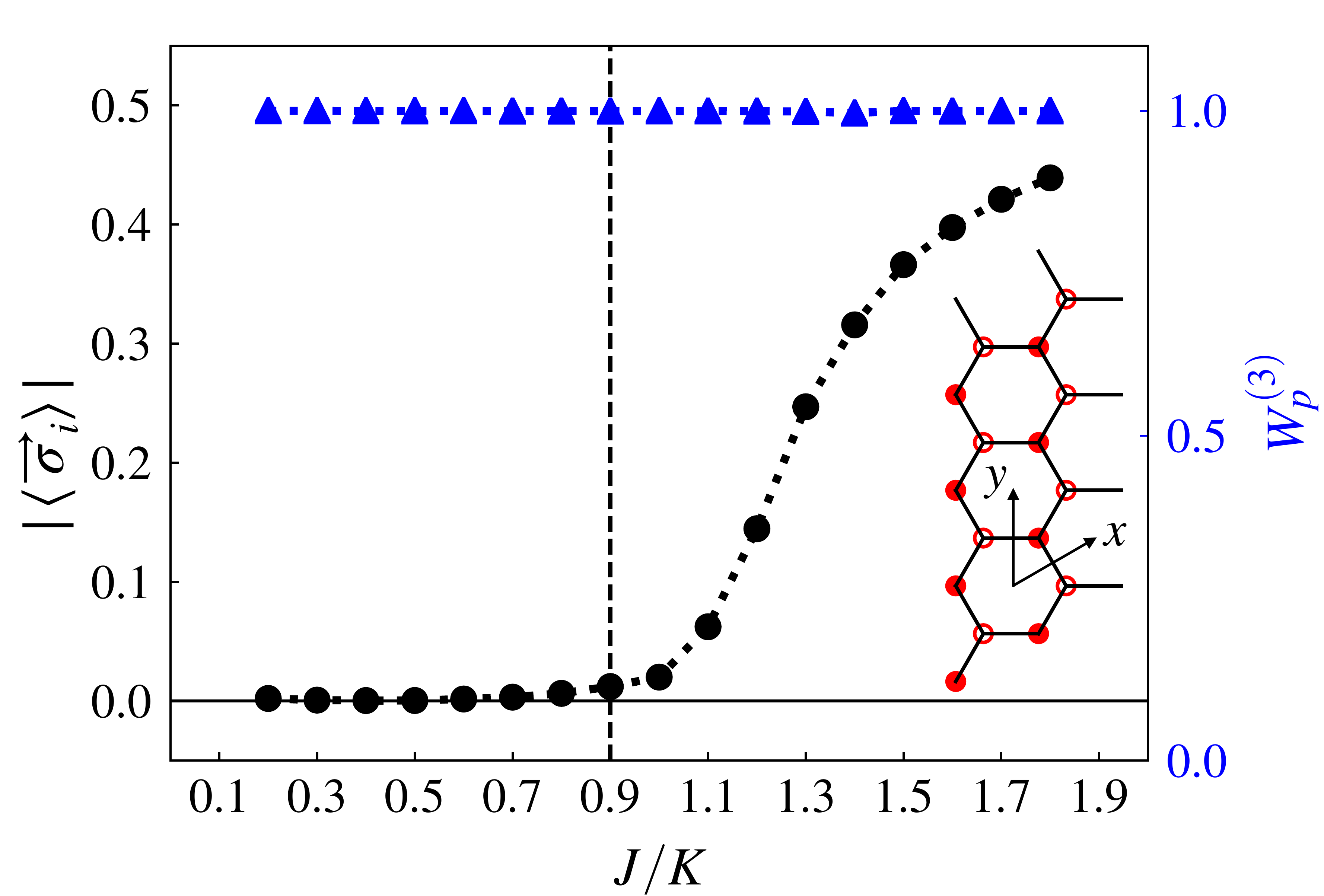}
\caption{The N\'{e}el spin-order parameter (black dots) and the ground-state expectation value of the plaquette operator $\Whex_p$ (blue triangles) as a function of $J/K$ in the spin-orbital model $\Hhex$ from iDMRG calculations performed on an infinite cylinder with circumference of $L_y = 4$ unit cells. The iDMRG uses two unit cells along the $x$-direction as its translationally invariant building block and the bond dimension is $\chi=1000$.}
\label{fig:DMRG}
\end{figure}

To validate the qualitative mean-field picture more quantitatively, we perform infinite density renormalization group (iDMRG) simulations \cite{schollwoeck2011,tenpy,dong2019} for the spin-orbital model $\Hhex$.
The results for the N\'{e}el spin-order parameter and the ground-state expectation value of the plaquette operator $\Whex_p$ are shown in Fig.~\ref{fig:DMRG}.
For the full range of $J/K$, the ground state stays in the zero-flux sector with $\Whex_p = 1$.
Furthermore, the N\'{e}el spin-order parameter shows a direct continuous transition from the paramagnetic $\Cm=3$ spin-orbital liquid to the N\'{e}el spin-ordered state at $J_\mathrm{c} \approx 0.9K$.
The critical coupling is larger than in the mean-field approximation, as expected.

{\em Gross-Neveu-$\text{SO}(3)$* criticality.}---%
Using a gradient expansion of the Majorana lattice model, we derive a continuum field theory describing the quantum critical point in the lowest flux sector.
To this end, we fix the gauge $u_{ij} = +1$ for $i\in A,j \in B$ sublattice and introduce continuum complex fermion fields $\psi^{\alpha}_s(\bvec x)$ with sublattice index $s=A,B$ and flavor index $\alpha = x,y,z$ by expanding the lattice Majorana fermions $c_{i,s}^\alpha = \psi_s^\alpha(\bvec x)\,\eu^{\iu \bvec{K} \cdot \bvec x} + \hc$ about the single Dirac point~$\bvec K$.
Upon a Hubbard-Stratonovich decoupling of the resulting four-fermion term $[\bar\psi(\mathds{1}_2 \otimes \vec L)\psi]^2 \mapsto \frac12 \vec \varphi^2 - \vec \varphi \cdot \bar\psi(\mathds{1}_2 \otimes \vec L)\psi$, with $\psi \equiv (\psi^x_A, \psi^y_A, \psi^z_A, \psi^x_B, \psi^y_B, \psi^z_B)^\top$, $\bar\psi \equiv \psi^\dagger (\sigma^z \otimes \mathds{1}_3)$, and $\vec\varphi$ the continuum N\'{e}el order parameter field, we obtain the Euclidean action
\begin{align} \label{eq:S_crit}
    \mathcal{S} = \int \du^2 \bvec x \, \du \tau \,\Bigl[&\bar{\psi} \gamma^\mu \partial_\mu \psi + g \vec \varphi \cdot \bar \psi (\mathds{1}_2 \otimes \vec L) \psi \nonumber\\ &
    + \frac{1}{2} \vec\varphi \left(-\partial_\mu^2 + m^2 \right) \vec \varphi + \lambda \left(\vec \varphi \cdot \vec \varphi \right)^2 \Bigr].
\end{align}
Here, $(\gamma^\mu) = (\sigma^z,\sigma^x,\sigma^y) \otimes \mathds{1}_{3}$, $\mu=0,1,2$, corresponds to a six-dimensional representation of the Clifford algebra, the Pauli matrices $\vec\sigma$ [$\SO(3)$ generators $\vec L$] act on the sublattice (flavor) indices, and we have allowed for dynamics and the symmetry-allowed quartic self-interaction of the order parameter.
The theory describes three flavors of two-component Dirac fermions coupled to a vector order parameter that transforms in the fundamental representation of $\SO(3)$, corresponding to spin $1$.
The boson mass $m^2$ can be used to tune through the $\SO(3)$-symmetry-breaking transition. 
For $m^2>0$, the order parameter $\vec\varphi$ is gapped and can be integrated out, corresponding to the paramagnetic spin-orbital-liquid phase characterized by three gapless Dirac fermions. For $m^2<0$, the minimum of the potential occurs at finite $\vec\varphi$, indicating spontaneous $\SO(3)$ symmetry breaking and an interaction-induced band gap for two out of the three fermions, with the third one remaining gapless, corresponding to the N\'{e}el-spin-ordered phase.

The existence of a quantum critical point at vanishing renormalized mass $m^2=0$ can be shown using a standard $\epsilon$ expansion about the upper critical space-time dimension of four. 
The flow equations admit an infrared stable fixed point at $g_*^2 > 0$ and $\lambda_* > 0$ that is characterized by a set of universal exponents [\onlinecite{suppl}\textcolor{blue}{(e)}]. Extrapolating the one-loop results to $\epsilon = 1$, we obtain the estimates $\eta_\phi \approx 0.33$ and $1/\nu \approx 1.1$. 
The remaining exponents $\alpha$, $\beta$, $\gamma$, and $\delta$ can then be obtained by means of the usual hyperscaling relations \cite{herbut07}, and the dynamical critical exponent is $z=1$.
For completeness, we also quote the fermion anomalous dimension, which is accessible in the ordinary Gross-Neveu-$\SO(3)$ universality class only, reading $\eta_\psi \approx 0.17$.
Note that this Gross-Neveu-$\SO(3)$ critical point defines a new universality class different from those of the Gross-Neveu-$\SUtwo$ (=chiral Heisenberg) model \cite{rosenstein93, herbut09b, janssen14, zerf17, gracey18, knorr18, toldin15, otsuka16, buividovich18, lang19, liu19}, in which case the fermion mass term transforms in the fundamental representation of $\SUtwo$, corresponding to spin $1/2$, and the spectrum in the ordered phase is fully gapped.

Upon generalizing the theory to $N$ flavors of Dirac fermions coupled to an $\SO(3)$ vector order parameter, the critical properties can alternatively be computed within a $1/N$ expansion in fixed dimension.
Extrapolating the leading-order results [\onlinecite{suppl}\textcolor{blue}{(f)}] to the physical $N=3$, we obtain the estimates $\eta_\phi \approx 0.32$, $\eta_\psi \approx 0.14$, and $1/\nu \approx 0.5$. While the anomalous dimensions agree with the $\epsilon$-expansion results within $\sim$5\% accuracy, the agreement in the case of $1/\nu$ is significantly less favorable. This may be due to a sizable $1/N^2$ correction to this observable, similar to the situation in the Gross-Neveu-$\SUtwo$ model \cite{gracey18}.

{\em Discussion.}---%
We have studied novel transitions between topological phases with concomitant symmetry breaking by making use of unbiased numerical results and controlled analytical approaches. The quantum critical points that we have discovered feature gapless Majorana fermions coupled to gapped $\Ztwo$ gauge fields as well as gauge-invariant order-parameter bosons, and fall into a previously unknown family of fractionalized fermionic universality classes.
They represent controlled instances of the larger class of unconventional quantum phase transitions that are characterized by fractionalized excitations, which includes deconfined quantum critical points between different conventionally-ordered phases \cite{senthil04b, nahum15a, shao16, wang17, sato17}, between conventional and deconfined phases \cite{ghaemi06, assaad16, gazit18, janssen20, zhang20}, as well as between different deconfined phases \cite{metlitski15, janssen17, boyack18}. The models studied in this work belong to the latter class, with the deconfined modes at the critical point being in a one-to-one correspondence with the deconfined elementary excitations of the adjacent phases.

Our findings call for more detailed theoretical investigations of the Gross-Neveu* criticalities, in particular in the $\SO(3)$ case, for which currently only leading-order estimates are available. It would also be interesting to study the universal finite-size spectra, e.g., on the torus \cite{whitsitt16, schuler16, whitsitt17, schuler19}.

Our results may be relevant for $4d^1$ or $5d^1$ Mott insulators \cite{balents10,pereira17,natori19}.
For instance, $\alpha$-ZrCl$_3$ realizes strongly bond-dependent interactions in analogy to the $d^5$ Kitaev materials \cite{jackeli09}, and has been proposed as a candidate for an $\mathrm{SU}(4)$-symmetric spin-orbital liquid on the honeycomb lattice \cite{jackeli18}.
Similarly, in double perovskite systems, strong spin-orbit coupling can lead to $j_\text{eff}=3/2$ multiplets subject to bond-dependent exchange interactions \cite{balents17}. In particular, absence of ordering down to low temperatures has been observed in Ba$_2$YMoO$_6$, and Kitaev-type spin-orbital liquids have been proposed as candidate ground states \cite{natori16}.
In the above materials, resonant inelastic X-ray scattering and neutron scattering can separately probe spin and spin-orbital excitations, allowing to resolve potential partial order that we find in our models.
Kugel-Khomskii-type models with anisotropic exchange interactions have also been proposed to describe correlated insulating phases in twisted bilayers \cite{balents18,natori19,vish19,rubio20}. In this regard, it may be of interest to consider twisted bilayer configurations of Kitaev materials, such as $\alpha$-RuCl$_3$ \cite{trebst17, janssen19}.

In real materials, additional perturbations that generate fluctuations of the gauge field are present. When their magnitudes become of the order of the (unperturbed) flux gap, further transitions that might lead to confinement of the fractionalized particles can occur.
The study of such transitions represents another interesting direction for future research.

\begin{acknowledgments}{\em Acknowledgments.}---%
We thank John Gracey, Wilhelm Kr\"u\-ger, Shouryya Ray, and Carsten Timm for illuminating discussions and collaborations on related topics.
This work has been supported by the Deutsche Forschungsgemeinschaft (DFG) through SFB 1143 (project id 247310070) and the W\"{u}rzburg-Dresden Cluster of Excellence {\it ct.qmat} (EXC 2147, project id 390858490). X.Y.D. is supported by the European Research Council under the grant ERQUAF (715861). S.C.\ acknowledges funding by the IMPRS for Many Particle Systems in Structured Environment at MPI-PKS. The work of L.J.\ is funded by the DFG through the Emmy Noether program (JA2306/4-1, project id 411750675).\end{acknowledgments}

\bibliography{solcritical_main_v2}
\bibliographystyle{shortapsrev4-2}

\end{document}


\title{%
\texorpdfstring{Supplemental Material:\\ Fractionalized fermionic quantum criticality in spin-orbital Mott insulators}%
{Supplemental Material: Fractionalized fermionic quantum criticality in spin-orbital Mott insulators}
}

\author{Urban F. P. Seifert}
\affiliation{Institut f\"ur Theoretische Physik and W\"urzburg-Dresden Cluster of Excellence ct.qmat, Technische Universit\"at Dresden, 01062 Dresden, Germany}
\author{Xiao-Yu Dong}
\affiliation{Department of Physics and Astronomy, Ghent University, Krijgslaan 281, 9000 Gent, Belgium}
\author{Sreejith Chulliparambil}
\affiliation{Institut f\"ur Theoretische Physik and W\"urzburg-Dresden Cluster of Excellence ct.qmat, Technische Universit\"at Dresden, 01062 Dresden, Germany}
\affiliation{Max-Planck-Institut f{\"u}r Physik komplexer Systeme, N{\"o}thnitzer Stra{\ss}e 38, 01187 Dresden, Germany}
\author{Matthias Vojta}
\affiliation{Institut f\"ur Theoretische Physik and W\"urzburg-Dresden Cluster of Excellence ct.qmat, Technische Universit\"at Dresden, 01062 Dresden, Germany}
\author{Hong-Hao Tu}
\affiliation{Institut f\"ur Theoretische Physik and W\"urzburg-Dresden Cluster of Excellence ct.qmat, Technische Universit\"at Dresden, 01062 Dresden, Germany}
\author{Lukas Janssen}
\affiliation{Institut f\"ur Theoretische Physik and W\"urzburg-Dresden Cluster of Excellence ct.qmat, Technische Universit\"at Dresden, 01062 Dresden, Germany}

\begin{abstract}
This Supplemental Material provides details of the predicted spectral structure of the Gross-Neveu* universality classes, the Majorana representation of the spin-orbital operators, the strong-coupling limits of the models, the Majorana mean-field theory, as well as the $\epsilon$- and large-$N$ calculations.
\end{abstract}

\date{\today}

\maketitle
\tableofcontents

\setcounter{figure}{0}
\setcounter{equation}{0}
\renewcommand\thefigure{S\arabic{figure}}
\renewcommand\theequation{S\arabic{equation}}

\section{Predictions for finite-size spectra in the Gross-Neveu* universality classes}

\subsection{Finite-size torus spectroscopy and fractionalized quantum criticality}

The difference between a fractionalized universality class and its ordinary counterpart reveals itself most prominently in the structure of the energy spectrum at the quantum critical point.
This can be made explicit within a tower-of-states analysis, i.e., by computing the low-energy eigenstates and their quantum numbers of a \emph{finite} system, in which the ground-state manifold splits into distinct energy levels $E_i$, $i=1,\dots$.
At a quantum critical point with dynamical exponent $z=1$, these energies scale for large linear system size $L$ as $E_i = \xi_i c /L$ with \emph{universal} constants of proportionality $\xi_i$ and a nonuniversal, but level-independent, global velocity $c$.
Within a radial-quantization geometry, the numbers $\xi_i$ can be associated with the universal scaling dimensions of the corresponding conformal field theory~\cite{cardy85}.
On torus geometries, which are more easily accessible to numerical simulations above (1+1) dimensions, this operator-state correspondence is not known to hold anymore exactly; however, the numbers $\xi_i$ are still universal and in fact, at least for small energies, appear to resemble the spectrum of scaling dimensions quite closely~\cite{schuler16, whitsitt17}.
Moreover, these universal numbers provide a unique fingerprint of the corresponding universality classes and, in particular, allow one to distinguish fractionalized quantum critical points from their ordinary counterparts~\cite{whitsitt16}.

To set the stage, let us review the situation for the case of the purely bosonic Ising and Ising* transitions on the torus. The latter can be realized, for instance, in a toric code model perturbed by a longitudinal field~\cite{trebst07, schuler16}. 
While states in the Ising spectrum that are even under the $\Ztwo$ symmetry also appear in the Ising* spectrum, and at the same energy (to leading order in $1/L$), $\Ztwo$-odd states are missing completely in the Ising* spectrum. 
This ``selection rule'' can be understood as a consequence of the gauge invariance in the Ising* case, which forbids the $\Ztwo$-odd sector in the physical spectrum.
On the other hand, the Ising* spectrum consists of additional states, in particular ones that are located at very low energies.
They arise from the fact that for fixed periodic boundary conditions of the physical degrees of freedom (spins), four different boundary conditions of the parton fields (spinons), periodic or antiperiodic along the two spatial directions, become simultaneously possible. These states, together with the ground-state level, are distinguished by different eigenvalues $\pm 1$ of the Wilson loops winding around the torus and can be associated with the topological degeneracy of the adjacent topological phase in the thermodynamic limit.

\subsection{Gross-Neveu and Gross-Neveu* universality classes}

The finite-size torus spectra in the ordinary Gross-Neveu-$\Ztwo$ universality class have recently been studied within two models of interacting spinless fermions on honeycomb and $\pi$-flux lattices, which both feature a quantum critical point between a semimetallic Dirac phase and a $\Ztwo$-ordered massive phase~\cite{schuler19}.
For simplicity, let us assume in the following a finite-size momentum discretization that includes the Dirac points.
At the critical point, the ground-state level in the finite-size spectra is then twofold (quasi-)degenerate and consists of the vacuum level $1_T$ and a level $\sigma_T$ that can be associated with the ``Semenoff'' mass generated by breaking the sublattice symmetry~\cite{herbut09a}.
The first excited level is an eightfold degenerate multiplet associated with single-fermion excitations $\psi_T$, whereas the next level consists of particle-hole excitations that can be associated with ``Haldane'' and ``Kekul\'{e}'' instabilities, as well as the lowest-lying two-fermion excitations $(2\psi)_T$~\cite{schuler19}.

In the Gross-Neveu-$\Ztwo$* universality class, single-fermion excitations are not gauge invariant and the corresponding levels therefore cannot occur in the spectrum.
However, in contrast to the fractionalized bosonic transitions~\cite{schuler16}, the level $\sigma_T$ is not gauge forbidden and will occur in the Gross-Neveu-$\Ztwo$* spectrum.
Due to the topological character of the $\Cm=2$ spin-orbital liquid~\cite{CSVJT20}, we expect at very low energy additional twofold-degenerate and nondegenerate levels $1_{T'}$ and $1_{T''}$, respectively, which are distinguished by different eigenvalues of the Wilson loop operators winding around the torus along the two spatial directions. Similar copies $\sigma_{T'}$ and $\sigma_{T''}$ of the $\sigma_T$ level should also occur in the respective topological sectors.

For the Gross-Neveu-$\SO(3)$ universality classes, the finite-size torus spectra are not even known in the ordinary case. However, by analogy, we also expect low-lying single-fermion excitations occuring in the Gross-Neveu-$\SO(3)$ spectrum, which are forbidden in the Gross-Neveu-$\SO(3)$* spectrum.
Moreover, the Gross-Neveu-$\SO(3)$* spectrum should feature additional very-low-lying states corresponding to insertions of nontrivial $\Ztwo$ fluxes threading the torus.
As the order parameter transforms as a vector under $\SO(3)$, the corresponding $\sigma_{T}$ level should exhibit, in analogy to the Wilson-Fisher $\mathrm{O}(3)$ criticality~\cite{whitsitt17}, a threefold degeneracy already in the ordinary Gross-Neveu-$\SO(3)$ spectrum, and its fractionalized Gross-Neveu-$\SO(3)$* counterpart should feature additional topological copies.

\section{Spin-orbital basis and Majorana fermion representation}

The exactly solvable square- and honeycomb-lattice models considered here are the $\Cm = 2$ and $\Cm=3$ members of the family of generalized Kitaev models with $\Cm$ flavors of itinerant fermions \cite{CSVJT20}, given in terms of $2^{q+1}$-dimensional Gamma matrices with $q=\lfloor \Cm / 2 \rfloor$.
The four-dimensional Gamma matrices required for the definition of the $\Cm = 2$ and $\Cm = 3$ models can be expressed in terms of two sets of Pauli matrices as
\begin{equation}
    \Gamma^\alpha = -\sigma^y \otimes \tau^\alpha, \quad \Gamma^4 = \sigma^x \otimes \mathds{1}_2, \quad \text{and} \quad \Gamma^5 = -\sigma^z \otimes \mathds{1}_2 ,
\end{equation}
where the $\sigma^\alpha$ denote three Pauli matrices acting on a local spin-1/2 degree of freedom and the $\tau^\alpha$ Pauli matrices correspond to pseudospins acting on a twofold degenerate orbital.
The exact solution~\cite{CSVJT20} proceeds by introducing Majorana fermions $c,b^\mu$ with $\mu = 1,\dots,5$ to represent the Gamma matrices as
\begin{equation}
    \Gamma^{\mu} = \iu b^\mu c \quad\text{and} \quad \Gamma^{\mu \nu} = \iu b^\mu b^\nu,
\end{equation}
with $\Gamma^{\mu \nu} = \iu [\Gamma^\mu, \Gamma^\nu]/2$.
On the square (honeycomb) lattice, the $\Ztwo$ gauge field on the four (three) distinct links is formed by $u_{\langle ij\rangle_\gamma} = \iu b_i^\gamma b_j^\gamma$ with $\gamma = 1, \dots, 3 (4)$.
The remaining Majoranas $c, b^5$ (and $b^4$ on the honeycomb lattice) disperse throughout the lattice.
It is convenient to relabel $b^5 \to c^x$, $c \to c^y$ and further on the honeycomb lattice $b^4 \to c^z$ such that the generator of the $\SO(2)$ symmetry in the square lattice model is given by the $z$-component of the spin degree of freedom, $\sigma^z = - \iu c^x c^y$, which commutes with $\Hsq_K$.
In the $\SO(3)$-symmetric model on the honeycomb lattice, the generators of the spin rotation symmetry are given by the three spin Pauli matrices which in the above Majorana representation read after relabeling
\begin{equation} \label{eq:sigma_maj}
    \sigma^\alpha \otimes \mathds{1}_2 \mapsto -\iu \epsilon^{\alpha \beta \gamma} c^\beta c^\gamma / 2 \equiv \frac{1}{2} c^\top L^\alpha c,
\end{equation}
where we have defined the $3\times 3$ matrices $(L^\alpha)_{\beta \gamma} = - \iu \epsilon^{\alpha \beta \gamma}$, which are $\SO(3)$ generators in the fundamental representation, and abbreviated $c = (c^x,c^y,c^z)^\top$.
This Majorana representation of spins, Eq.~\eqref{eq:sigma_maj}, has been previsouly used to study pure spin models, see e.g., Refs.~\cite{tsvelik92,shastrysen97}.
Note that $\SO(3)$ rotations of the Majorana three-vector $c^\alpha \mapsto O^{\alpha \sigma} c^\sigma$ induce spin rotations as
\begin{equation}
    \sigma^\alpha \otimes \mathds{1}_2 \mapsto -\iu \epsilon^{\alpha \beta \gamma} O^{\beta \kappa} O^{\gamma \rho} c^\kappa c^\rho / 2 = O^{\alpha \lambda} \sigma^\lambda \otimes \mathds{1}_2,
\end{equation}
where we have inserted $\delta^{\sigma \alpha} = O^{\sigma \lambda} O^{\alpha \lambda}$ and subsequently used that $O^{\sigma \lambda} O^{\beta \kappa} O^{\gamma \rho} \epsilon^{\sigma \beta \gamma} \epsilon^{\lambda \rho \gamma} = \det O$ with $\det O = +1$ for $O \in \SO(3)$.

\section{Spin-orbital models in the strong-coupling limit}
In this section, we discuss the strong-coupling limits of the square-lattice model $\mathcal H^{(2)} = H_K^{(2)} + H_{J^z}^{(2)}$ for $|J^z/K| \to \infty$ and the honeycomb-lattice model $\mathcal H^{(3)} = H_K^{(3)} + H_{J}^{(3)}$ for $|J/K| \to \infty$.
These can be understood within perturbative expansions in small $K/J^z$ and $K/J$, respectively.
Note that both $H_{J^z}^{(2)}$ and $H_{J}^{(3)}$ commute with all operators that act only on the orbital degrees of freedom.
The ground state with long-range order in the spin sector in both model therefore features a macroscopic degeneracy in the orbital sector for $K=0$.

Switching on an infinitesimal $K > 0$ for antiferromagnetic $J^z > 0$ in the square-lattice model $H^{(2)}$, the orbital degeneracy is lifted in fourth-order perturbation theory in $K/J^z$ via a process that flips spins along a plaquette pairwise, giving rise to an effective $\Ztwo$ gauge theory in the orbital sector. 
This is in contrast to the situation for ferromagnetic $J^z < 0$, in which case the spin-polarized ground state of $H_{J^z}^{(2)}$ is an eigenstate of $\mathcal{H}^{(2)}_K$ with zero energy, so that the degeneracy is not lifted at any order in perturbation theory.
Below a certain finite critical $J^z/K < 0$, there is therefore an extended phase in which the orbital sector features a macroscopic ground-state degeneracy, similar to the situation in a strong magnetic field for $J^z = 0$~\cite{CJVTS20}.
In the honeycomb-lattice model $H^{(3)}$, a small $K/J$ lifts the orbital degeneracy already at first order in $K/J$, independent of the sign of~$J$. 

We therefore expect that the long-range-ordered phases in both models at $J^z/K > (J^z/K)_\mathrm{c}>0$ and $J/K > (J/K)_\mathrm{c}>0$ extend all the way up to $J^z \to \infty$ and $J \to \infty$, without any additional phase transition at finite coupling.

\section{Majorana mean-field theory for the honeycomb-lattice model}

\subsection{Mean-field decoupling and solution}

The Hamiltonian for the $\Cm = 3$ spin-orbital liquid with a spin-only Heisenberg interaction reads
\begin{equation}
    \Hhex = - K \sum_{\langle ij \rangle} \vec \sigma_i \cdot \vec \sigma_j \otimes \tau^\alpha_i \tau^\alpha_j + J \sum_{\langle ij \rangle} \vec \sigma_i \cdot \vec \sigma_j \otimes \mathds{1}.
\end{equation}
Proceeding with the Majorana representation, one obtains 
\begin{subequations}
\begin{align}\label{eq:k-j-maj}
    \Hhex & \mapsto K \sum_{\langle ij \rangle} \iu u_{ij} c^\alpha_i c^\alpha_j + J \sum_{\langle ij \rangle} \left( \frac{1}{2} c^\top_i L^\alpha c_i \right) \left( \frac{1}{2} c_j^\top L^\alpha c_j \right) \\
    &= K \sum_{\langle ij \rangle} \iu u_{ij} c^\alpha_i c^\alpha_j + J \sum_{\langle ij \rangle} \left( \frac{1}{2} c^\beta_i c^\beta_j c^\gamma_i c^\gamma_j + \frac{3}{2} \right) \label{eq:k-j-maj-2}
    \end{align}
\end{subequations}
We decouple the interaction in Eq.~\eqref{eq:k-j-maj} into onsite mean fields $\sigma_i^\alpha \simeq \langle \sigma^\alpha_i \rangle = \langle c^\top_i L^\alpha c_i / 2 \rangle \in \mathbb{R}$ and bond variables $\iu c^\alpha_i c^\alpha_j \simeq \langle \iu c^\alpha_i c^\alpha_j \rangle \equiv \chi^\alpha_{ij} \in \mathbb{R}$ (no summation convention),
\begin{align}
    \left( \frac{1}{2} c^\top_i L^\alpha c_i \right) \left( \frac{1}{2} c_j^\top L^\alpha c_j \right) 
    & \simeq 
    \langle \sigma^\alpha_i \rangle  \left( \frac{1}{2} c_j^\top L^\alpha c_j \right) +  \langle \sigma^\alpha_j \rangle \left( \frac{1}{2} c_j^\top L^\alpha c_j \right) - \langle \sigma_i^\alpha \rangle \langle \sigma_j^\alpha \rangle 
    - \sum_{\beta,\gamma} |\epsilon^{\alpha \beta \gamma}| \left( \chi^\beta_{ij} \iu c^\gamma_i c^\gamma_j - \frac{1}{2} \chi_{ij}^\beta \chi_{ij}^\gamma \right).
\end{align}
While $\langle \sigma^\alpha \rangle$ transforms as a vector ($\mathbf{3}$ representation) under $\SO(3)$, the $\chi^\alpha_{ij}$ are the diagonal components of the rank-2 mean-field tensor $\langle \iu c^\alpha c^\beta \rangle$.
As the interaction does not warrant decoupling into off-diagonal components, the bond variables can only transform in the $\mathbf{0}$ irreducible representation, which is given by the trace of above tensor. We therefore in the following restrict the mean-field parameter as $\chi^\alpha_{ij} \equiv \chi_{ij} = \sum_\beta \langle \iu c^\beta_i c^\beta_j \rangle / 3$ for all $\alpha=x,y,z$.
This result is also straightforwardly obtained by directly decoupling the second term in Eq.~\eqref{eq:k-j-maj-2} in a manifestly $\SO(3)$-invariant manner (i.e., by choosing a decoupling which respects the summation convention).

\begin{figure}
    \centering
    \includegraphics[width=0.6\textwidth]{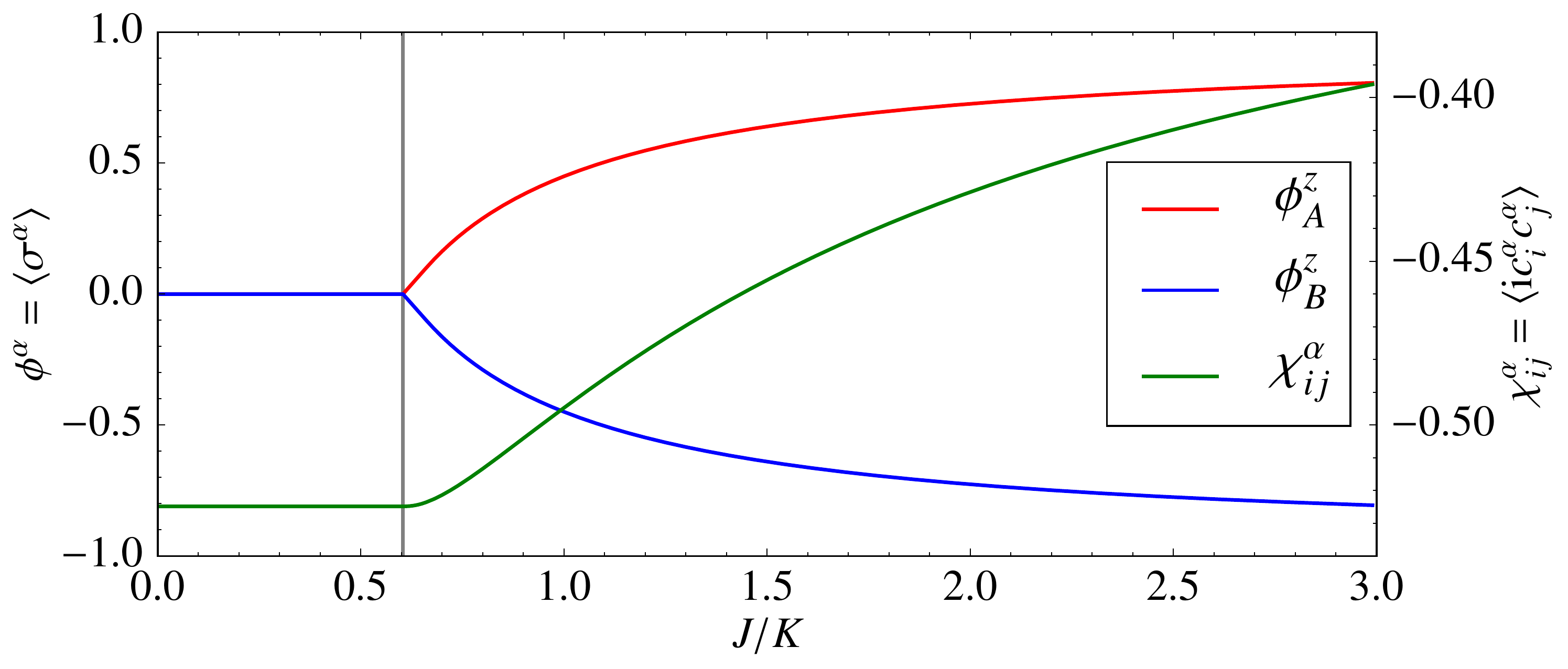}
    \caption{Spin mean fields $\phi_A^z$ and $\phi_B^z$ (red and blue lines, corresponding to the two sublattices $A$ and $B$) and bond variable $\chi^\alpha_{ij}$ (green line) in the honeycomb-lattice model $\Hhex$ from Majorana mean-field theory at temperature $T=0.05K$, using a lattice with $N_\text{uc} = 48\times48$ unit cells. Here, we have assumed homogeneous and flavor-independent bond variables $\chi^\alpha_{ij} \equiv \chi$ and the mean fields $\langle \sigma^\alpha_i \rangle \equiv \phi_A^\alpha \text{ or } \phi_B^\alpha$ to only depend on the sublattice for given $J/K$. The critical value of $J_c/K\approx 0.604$ (vertical grey line) is obtained by a linear fit of $\phi
   ^z_A$ as a function of $J/K$ near the critical point.}
    \label{fig:mft}
\end{figure}

Working in the flux-free sector, choosing the gauge $u_{ij} = +1$ for $i \in A$ and $j \in B$ sublattices is convenient. The Majorana hopping problem then possesses the translational and rotational symmetries of the honeycomb lattice, such that we assume $\chi_{ij} \equiv \chi$ to be homogeneous and $\langle \sigma_i^\alpha \rangle$ to only depend on the sublattice for given $J/K$, totalling ($1 + 3 \times 2$) mean field parameters.
We employ a Fourier decomposition of the Majorana fermions as
\begin{equation}
    c_{i,s}^\alpha = \sqrt{\frac{2}{N_\text{uc}}} \sum_{\bvec{k} \in \mathrm{BZ}/2} \left[ c^\alpha_{s,\bvec k} \eu^{\iu \bvec{k} \cdot \bvec{r_i}} + \left( c^\alpha_{s,\bvec{k}} \right)^\dagger \eu^{-\iu \bvec{k} \cdot \bvec{r_i}}  \right],
\end{equation}
with $N_\text{uc}$ denoting the number of unit cells, the sublattice index $s=A,B$, and the momentum summation is over half of the Brillouin zone (BZ). The mean-field Hamiltonian then reads
\begin{equation} \label{eq:mft_hamiltonian}
    \Hhex_\mathrm{MFT} = \sum_{\bvec{k} \in \mathrm{BZ}/2} c_{\bvec{k}}^\dagger \left[ \begin{pmatrix} 0 & \iu ( K - J \chi) f(\bvec k) \\
                    -\iu (K-J \chi) f^\ast(\bvec k) & 0
    \end{pmatrix} \otimes \mathds{1}_{3} + \begin{pmatrix}  6 J \langle \vec \sigma_B \rangle & 0 \\ 0 & 6 J \langle \vec \sigma_A \rangle \end{pmatrix} \cdot \vec L  \right] c_{\bvec{k}} - 3 J N_\text{uc} \left(\langle \vec \sigma_A \rangle \cdot \langle \vec \sigma_B \rangle + \chi^2 \right),
\end{equation}
where we have introduced the six-component spinor $c_{\bvec{k}} = (c_{\bvec{k},A}^x, c_{\bvec{k},A}^y, \dots,c_{\bvec{k},B}^z)^\top$ and $f(\bvec q) = 1+ \eu^{\iu \bvec q \cdot \bvec n_1} + \eu^{\iu \bvec q \cdot \bvec n_2}$, where $\bvec n_{1,2} = (\pm 1, \sqrt{3})/\sqrt{2}$ denote the reciprocal lattice vectors.
We then solve the self-consistency equations for the mean-field parameters iteratively.
A representative solution is depicted in Fig.~\ref{fig:mft}, showing a continuous evolution of the N\'eel spin-order parameter $n^z = (\langle \vec \sigma_A^z \rangle - \langle \vec \sigma_B^z \rangle ) / 2 = (\phi_A^z - \phi_B^z)/2$ above the quantum critical point at $J_\mathrm{c} \approx 0.604 K$.

\subsection{Spectrum in the symmetry-broken phase}

From Eq.~\eqref{eq:mft_hamiltonian}, it is apparent that one can straightforwardly diagonalize $\Hhex_\mathrm{MFT}$ in flavor space by some unitary $U$
\begin{equation} \label{eq:diag_L}
    (\mathds{1}_{2} \otimes U)^\dagger 
    \left[\begin{pmatrix} - 6 J \vec n & 0 \\ 0 & 6 J \vec n \end{pmatrix} \cdot \vec L \right]
    (\mathds{1}_{2} \otimes U) =  6 J |\vec n | \  \Sigma^z \otimes \diag( 1 , 0, - 1),
\end{equation}
where we have inserted the definition of the Néel order parameter $\vec n = (\langle \vec \sigma_A \rangle - \langle \vec \sigma_B \rangle ) / 2$ and $\Sigma^z$ denotes the third Pauli matrix acting on sublattice degrees of freedom.
Thus the Hamiltonian $\Hhex_\mathrm{MFT}$ is block-diagonal with three $2 \times 2$ blocks.
Since $\Sigma^z$ anticommutes with the first term in Eq.~\eqref{eq:mft_hamiltonian}, two of these blocks contain mass terms which gap out the respective Dirac nodes, with the dispersion reading
\begin{equation}
    \varepsilon_\pm^{(+1,-1)} (\bvec q) = \pm \sqrt{(K')^2 |f(\bvec q)|^2 + \left(6 J |\vec n|\right)^2}, \qquad \text{with $\varepsilon_\pm^{(+1,-1)} \gtrless 0$ for all $\bvec q \in \text{BZ}/2$},
\end{equation}
while the remaining block yields a gapless Dirac node with renormalized Kitaev coupling parameters $K' = K - J \chi$
\begin{equation}
    \varepsilon^{(0)}_\pm(\bvec q) = \pm | K' f(\bvec q)|, \qquad \text{with $\varepsilon_\pm^{(0)} = 0$ for $\bvec q = \mathbf{K} \in \text{BZ}/2$.}
\end{equation}

\section{Critical behavior of the Gross-Neveu-SO(3) model}

\subsection{\texorpdfstring{$4-\epsilon$ expansion}{4-epsilon expansion}}

\begin{figure}[t]
    \centering
    \includegraphics[width=\textwidth]{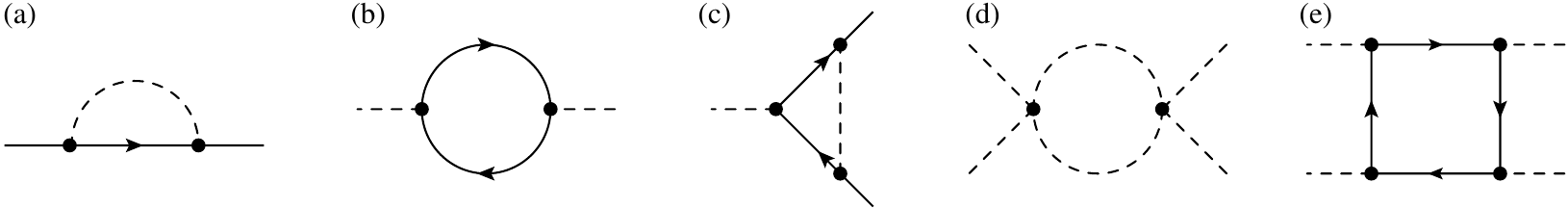}
    \caption{Feynman diagrams that contribute to the flow equations of the Gross-Neveu-$\SO(3)$ model at leading order in the $4-\epsilon$ expansion, showing the fermion (a) and boson (b) self energies, and the vertex corrections to the Yukawa coupling $g$ (c) and the scalar coupling $\lambda$ (d,e). Solid (dashed) inner lines correspond to fermion (boson) propagators.}
    \label{fig:diagrams-eps}
\end{figure}

The Gross-Neveu-$\SO(3)$ model defined in Eq.~\eqScrit\ of the main text admits a stable renormalization group fixed point in the critical plane $m^2 = 0$ that can be accessed within a standard $\epsilon$ expansion about the upper critical space-time dimension of four. 
At one-loop order, this amounts to evaluating the Feynman diagrams displayed in Fig.~\ref{fig:diagrams-eps}.
Integrating over the momentum shell between $\Lambda/b$ and $\Lambda$ with $b>1$ in $D=4-\epsilon$ space-time dimensions causes the coupling to flow as 
\begin{align}
    \frac{\du g^2}{\du \ln b} & = \epsilon g^2  - \frac{2(N+6)}{3} g^4,\\
    \frac{\du \lambda}{\du \ln b} & = \epsilon \lambda  - \frac{4N}{3}g\lambda - 44 \lambda^2 + \frac{N}{3}g^4,
\end{align}
where we have rescaled $g^2/(8\pi^2\Lambda^\epsilon) \mapsto g^2$ and $\lambda/(8\pi^2\Lambda^\epsilon) \mapsto \lambda$, and allowed for a general number $N$ of two-component Dirac fermions, with $N=3$ in the case relevant for the spin-orbital model.
The infrared stable fixed point is located at 
\begin{align}
    g_*^2 & = \frac{3}{2(N+6)}\epsilon + \mathcal O(\epsilon^2), \\ 
    \lambda_* & = \frac{6-N + \sqrt{N^2+120 N + 36}}{88(N+6)}\epsilon + \mathcal O(\epsilon^2),
\end{align}
and is characterized by the anomalous dimensions
\begin{align} \label{eq:eta-eps}
    \eta_\phi = \frac{N}{N+6}\epsilon + \mathcal O(\epsilon^2)
    \qquad \text{and} \qquad
    \eta_\psi = \frac{3}{2(N+6)}\epsilon + \mathcal O(\epsilon^2).
\end{align}
The correlation-length exponent can be computed from the flow of the tuning parameter in the vicinity of the critical point,
\begin{align}
    \frac{\du m^2}{\du \ln b} = 2 m^2 -\frac{2N}{3} g^2 m^2 + 20 \frac{\lambda}{1+m^2} - \frac{4N}{3}g^2,
\end{align}
from which we obtain
\begin{align} \label{eq:nu-eps}
    1/\nu = 2 - \frac{17N+30+5\sqrt{N^2+120N+36}}{22(N+6)}\epsilon + \mathcal{O}(\epsilon^2).
\end{align}
Setting $N=3$ and extrapolating to $\epsilon = 1$, we obtain the estimates for $\eta_\phi$, $\eta_\psi$ and $1/\nu$ as quoted in the main text.
For comparison with the large-$N$ calculation discussed below, it is also useful to consider the large-$N$ limits of Eqs.~\eqref{eq:eta-eps} and \eqref{eq:nu-eps}, reading
\begin{align} \label{eq:eps-largeN}
    \eta_\phi & = \epsilon - \frac{6\epsilon}{N} + \frac{36\epsilon}{N^2} + \mathcal O(\epsilon^2, 1/N^3), &
    \eta_\psi & = \frac{3\epsilon}{2N} - \frac{9\epsilon}{N^2} + \mathcal O(\epsilon^2, 1/N^3), &
    1/\nu & = 2 - \epsilon - \frac{9\epsilon}{N} + \frac{459\epsilon}{N^2} + \mathcal O(\epsilon^2, 1/N^3).
\end{align}

\subsection{\texorpdfstring{Large-$N$ expansion}{Large-N expansion}}

\begin{figure}[t]
    \centering
    \includegraphics[width=.85\textwidth]{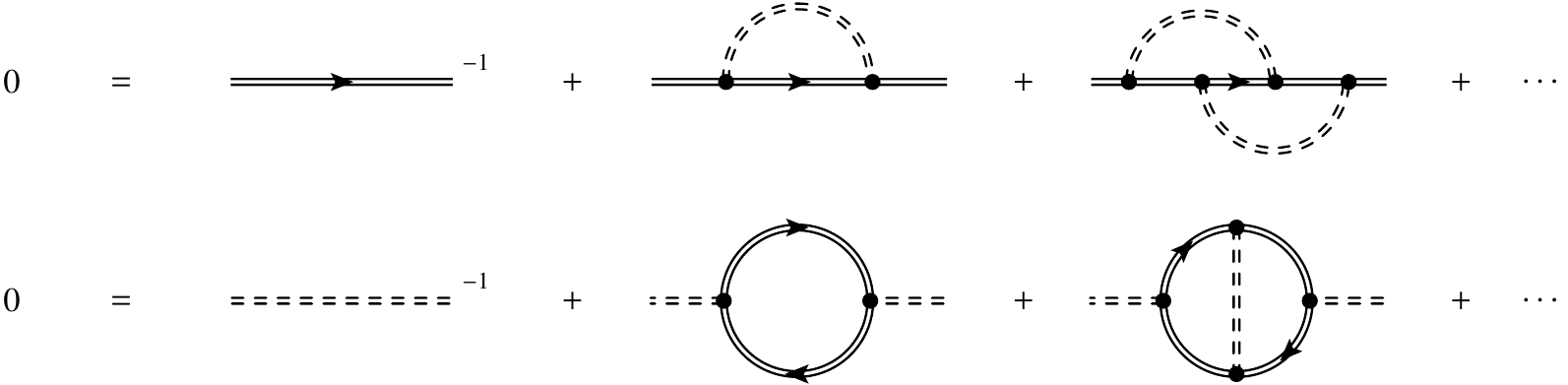}
    \caption{Skeleton Schwinger-Dyson equations used to determine the critical behavior of the Gross-Neveu-$\SO(3)$ model at leading order in the large-$N$ expansion. Double solid (dashed) inner lines correspond to the scaling forms of the fermion (boson) fields at criticality.}
    \label{fig:diagrams-largeN}
\end{figure}

The critical properties of the Gross-Neveu-$\SO(3)$ model can alternatively be computed within a large-$N$ expansion in fixed space-time dimension $D$. We employ the critical-point formalism~\cite{vasilev81, gracey91, graceyreview}, which allows us to deduce the analytical continuation of the critical exponents in fractional dimension $2<D<4$ efficiently.
This approach exploits the scale invariance of the system at criticality, thereby providing a means to systematically solve the Schwinger-Dyson equations order by order in $1/N$ algebraically. Specifically, we assume asymptotic scaling forms of the fermion and boson fields, reading in coordinate space
\begin{align}
    \langle\psi\bar\psi\rangle(x) & \sim \frac{A \slashed{x}}{\left(x^2\right)^{(D+\eta_\psi)/2}} \left[1 + A' \left(x^2\right)^{1/(2\nu)} \right],
    &
    \langle\vec\varphi\vec\varphi^\top\rangle(x) & \sim \frac{B}{\left(x^2\right)^{(D-2+\eta_\phi)/2}} \left[1 + B' \left(x^2\right)^{1/(2\nu)} \right]  \mathds{1}_3 ,
\end{align}
where $\slashed{x} \equiv \gamma^\mu x_\mu$ and $A$, $B$ and $A'$, $B'$ are $x$-independent amplitudes for the dominant scaling forms and corrections to scaling, respectively, as $x \to 0$ \cite{gracey18}.
Here, $(\gamma^\mu) = (\sigma^z, \sigma^x, \sigma^y) \otimes \mathds{1}_N$, $\mu = 0, 1, \dots, D-1$, corresponds to a $2N$-dimensional representation of the Clifford algebra, $\{\gamma^\mu, \gamma^\nu\} = 2 \delta^{\mu\nu} \mathds{1}_{2N}$, with $N$ being the number of two-component Dirac fermions, and we assume that $N$ is a multiple of three.
The skeleton Schwinger-Dyson equations for the two-point functions used to compute all three exponents $\eta_\phi$, $\eta_\psi$, and $1/\nu$ at order $\mathcal O(1/N)$ are displayed diagrammatically in Fig.~\ref{fig:diagrams-largeN}.
The explicit evaluation follows closely the analogous computation in the Gross-Neveu-SU(2) universality class \cite{gracey18}, and will therefore not be repeated here.
We find
\begin{align} \label{eq:largeN}
    \eta_\phi & = 4 - D + \frac{3(4-3D)}{D} \frac{C_D}{N} + \mathcal O(1/N^2),
    &
    \eta_\psi & = \frac{3(D-2)}{D} \frac{C_D}{N} + \mathcal O (1/N^2),
    &
    1/\nu & = (D-2)\left(1 - \frac{6(D-1)}{D} \frac{C_D}{N}\right) + \mathcal O(1/N^2),
\end{align}
where $C_D^{-1} \equiv \Gamma(2-{D}/{2})\Gamma({D}/{2})\Beta({D}/{2},{D}/{2}-1)$, with $\Gamma(\,\cdot\,)$ and $\Beta(\,\cdot\,,\,\cdot\,)$ denoting the gamma and beta functions, respectively.
In the physical case $D=3$, we obtain
\begin{align} \label{eq:largeN-D3}
    \eta_\phi & = 1 - \frac{20}{\pi^2 N} + \mathcal O(1/N^2), 
    &
    \eta_\psi & = \frac{4}{\pi^2 N}  + \mathcal O(1/N^2), 
    &
    1/\nu & = 1 - \frac{16}{\pi^2 N} + \mathcal O(1/N^2).
\end{align}
We have explicitly verified Eqs.~\eqref{eq:largeN-D3} using a conventional large-$N$ calculation analogous to Ref.~\cite{boyack19}.
The estimates for the transition in the spin-orbital model on the honeycomb lattice are obtained by extrapolating to $N=3$.

Upon expanding Eqs.~\eqref{eq:largeN} in $D=4-\epsilon$ dimensions, we find
\begin{align}
    \eta_\phi & = \epsilon - \frac{6\epsilon}{N} + \frac{15\epsilon^2}{4N} + \mathcal O(\epsilon^3, 1/N^2),
    &
    \eta_\psi & = \frac{3\epsilon}{2N} - \frac{9\epsilon^2}{8N} + \mathcal O(\epsilon^3,1/N^2),
    &
    1/\nu & = 2 - \epsilon - \frac{9\epsilon}{N} + \frac{39\epsilon^2}{4N} + \mathcal O(\epsilon^3,1/N^2),
\end{align}
which agrees precisely with Eqs.~\eqref{eq:eps-largeN} up to the respective orders calculated. This furnishes a nontrivial crosscheck of both the $\epsilon$-expansion and large-$N$ calculations.

\bibliography{solcritical_main_v2}
\bibliographystyle{shortapsrev4-2}